# Supergiant Fast X-ray Transients with Swift: catching outbursts and monitoring programs

P. Romano,[1] J.A. Kennea,[2] D.N. Burrows,[2] V. Mangano,[1] V. La Parola,[1] G. Cusumano,[1] S. Vercellone,[1] P. Esposito,[3] H.A. Krimm,[4] C. Pagani,[5] and N. Gehrels[4]

[1] INAF, Istituto di Astrofisica Spaziale e Fisica Cosmica,
Via U. La Malfa 153, I-90146 Palermo, Italy
[2] Department of Astronomy and Astrophysics, Pennsylvania State University,
University Park, PA 16802, USA
[3] INAF, Osservatorio Astronomico di Cagliari,
località Poggio dei Pini, strada 54, I-09012 Capoterra, Italy
[4] NASA/Goddard Space Flight Center, Greenbelt, MD 20771, USA
[5] Department of Physics & Astronomy, University of Leicester, LE1 7RH, UK
E-mail(PR): romano@ifc.inaf.it

ABSTRACT

*Swift* is shedding new light on the phenomenon of Supergiant Fast X–ray Transients (SFXTs), a recently discovered class of High-Mass X-ray Binaries, whose optical counterparts are O or B supergiants, and whose X–ray outbursts are about 10,000 times brighter than their quiescent state. Thanks to its unique automatic fast-slewing and broad-band energy coverage, *Swift* is the only observatory which can detect outbursts from SFXTs from the very beginning and observe their evolution panchromatically. Taking advantage of *Swift*'s scheduling flexibility, we have been able to regularly monitor a small sample of SFXTs with 2–3 observations per week (1–2 ks) for two years with the X–Ray Telescope (XRT). Our campaigns cover all phases of their lives, across 4 orders of magnitude in flux. We report on the most recent outburst of AX J1841.0−0536 caught by *Swift* which we followed in the X–rays for several days, and on our findings on the long-term properties of SFXTs and their duty cycle.

## 1. Introduction

Supergiant Fast X–ray Transients (SFXTs) are a subclass of High Mass X–ray Binaries (HMXBs) discovered by INTEGRAL (Sguera et al. 2005), associated with an O or B supergiant. They display outbursts which are characterized by bright flares with a duration of a few hours (as seen by INTEGRAL), a peak luminosity of $10^{36}$–$10^{37}$ erg s$^{-1}$ (Sguera et al. 2005, 2006; Negueruela et al. 2006), and a very large dynamic range (3–5 orders of magnitude), as their quiescence luminosity is $\sim 10^{32}$ erg s$^{-1}$ (in't Zand 2005). Their hard X–ray spectra have the typical shape of HMXBs hosting X–ray pulsars, a flat hard power law below 10 keV, with a high energy cut-off at about 15–30 keV, sometimes strongly absorbed at soft energies (Walter et al. 2006; Sidoli et al. 2006). Therefore, it is generally assumed that SFXTs are HMXBs hosting a neutron star (NS), even though pulse periods are only detected in 4 of the 10 confirmed members of the class. About 20 more candidates are known which showed short transient flaring activity, but which have no confirmed association with an OB supergiant companion. The field is rapidly evolving, so this number is likely to increase in the near future. Consensus has not yet been reached as to the nature of the mechanism producing the outbursts, which is probably related to the properties of the wind from the supergiant companion (in't Zand 2005; Walter & Zurita Heras 2007; Negueruela et al. 2008; Sidoli et al. 2007) or to the presence of a centrifugal or magnetic barrier (Grebenev & Sunyaev 2007; Bozzo et al. 2008).

These enigmatic sources have been recently observed by NASA's *Swift* Gamma-Ray Burst Explorer Mission (Gehrels et al. 2004), whose fast-slewing capability and broad-band energy coverage, paired with its flexible observing scheduling, make it an ideal facility to study both the bright outburst and the out-of-outburst behaviours.

Important information is also coming from "Monitor of the All-Sky X-ray Image" (MAXI, Matsuoka et al. 2009). MAXI recently observed an outburst of the SFXT AX J1841.0−0536 on 2010 November 7 (Negoro et al. 2010), during which the source was faintly detected for almost a day (with an average 0.5–20 keV flux of 20 mCrab), and a peak of activity that reached ∼ 60 mCrab. The spectrum could be fit by an absorbed power-law with a photon index of 0.9.

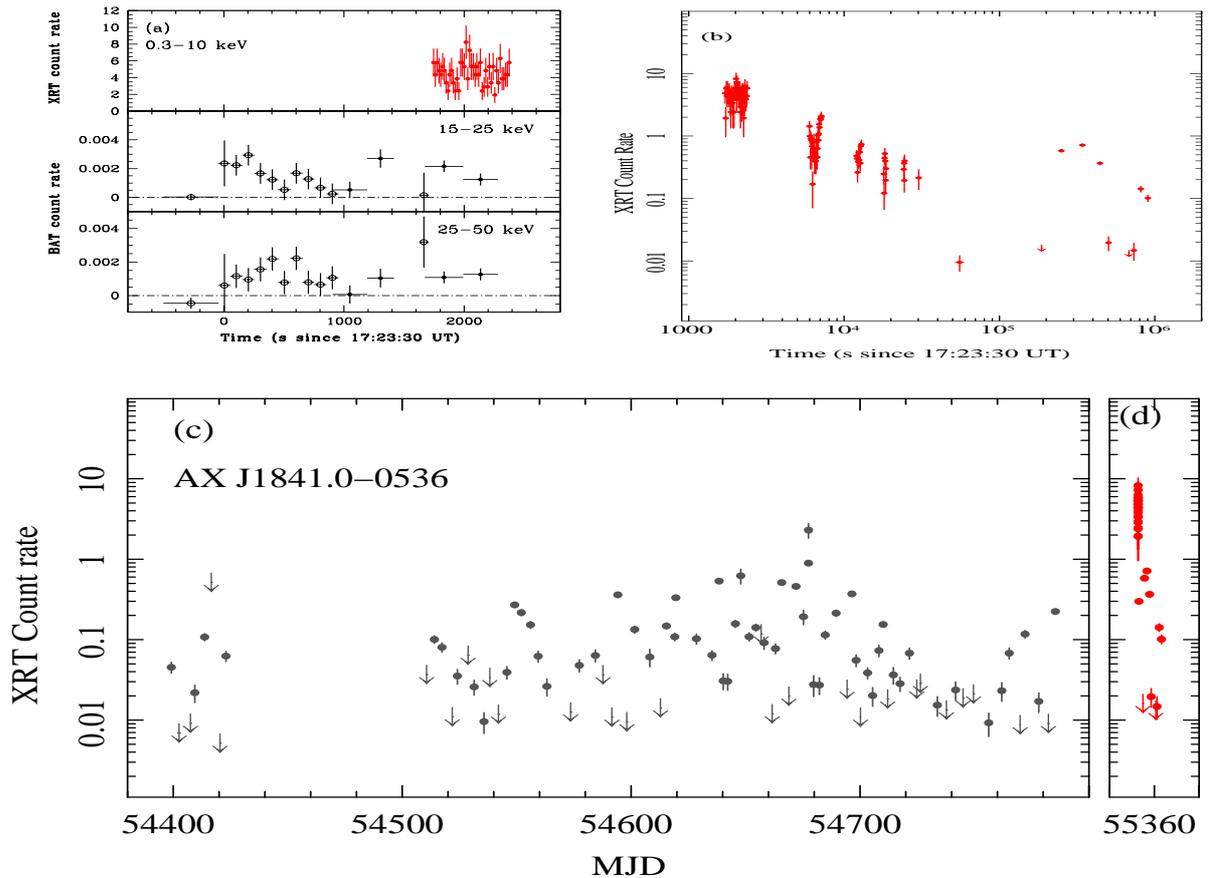

Fig. 1. Light curves of AX J1841.0−0536. (a): XRT (red) and BAT (black) light curves of the 2010 June 5 outburst in units of count s$^{-1}$ and count s$^{-1}$ detector$^{-1}$, respectively during the first orbit. The (black) empty circles correspond to BAT in event mode (S/N= 5), filled (black) circles to BAT survey mode data. The downward-pointing arrows are 3-$\sigma$ upper limits. (b): XRT complete follow-up of the outburst (0.3–10 keV). (c): XRT light curve (0.2–10 keV; grey) of the 2007–2008 monitoring campaign (Romano et al. 2009). (d): light curve of the 2010 June 5 outburst in the same time-scale. Adapted from Romano et al. (2011b).

In this paper we describe the *Swift* contribution to the SFXT science[1], by first reporting on the *Swift* observations of the 2010 June 5 outburst of AX J1841.0−0536— so far the only SFXT that was also detected by MAXI— and then on the *Swift* monitoring campaigns.

## 2. Catching Outbursts: the case of AX J1841.0−0536

In general, the spectrum of a SFXT can be described in terms of a hard power law below 10 keV with a high-energy cutoff at 15–30 keV. Simultaneous observations performed by the *Swift* X–ray Telescope (XRT, Burrows et al. 2005) and Burst Alert Telescope (BAT, Barthelmy et al. 2005) can provide broad-band spectroscopy of outbursts of SFXTs from 0.3 keV to 100–150 keV, thus constraining both the absorption and the hard X–ray spectral properties. In order to ensure simultaneous narrow field instrument data, the *Swift* Team enabled automatic rapid slews to the whole sample of SFXTs following detection of flares by the BAT, in the same way as is currently done for GRBs.

As an example, we report the properties of the 2010 June 5 outburst of AX J1841.0−0536 (Figure 1) for which simultaneous BAT and XRT data were collected. AX J1841.0−0536 was discovered during *ASCA* observations of the Scutum arm region performed in 1994 and 1999. It was seen to be a flaring source which showed flux increases by a factor of 10 (up to ∼ $10^{-10}$ erg cm$^{-2}$ s$^{-1}$) with rising times on the order of 1 hr (Bamba et al. 2001), with strong absorption ($N_{\rm H} = 3 \times 10^{22}$ cm$^{-2}$), and coherent pulsations with a period of 4.7394±0.0008 s. Halpern

---

[1] http://www.ifc.inaf.it/sfxt/

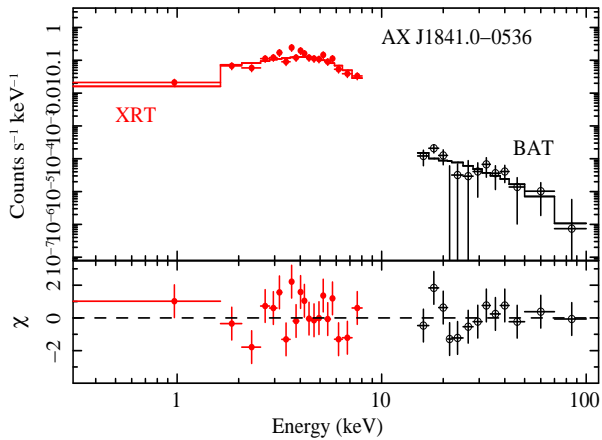

Fig. 2. Spectroscopy of the 2010 June 5 outburst of AX J1841.0−0536. **Top:** simultaneous XRT (filled red circles) and BAT (empty blue circles) data fit with the HIGHECUT model. **Bottom:** the residuals of the fit (in units of standard deviations). Data from Romano et al. (2011b).

& Gotthelf (2004) identified it with IGR J18410−0535, a newly discovered source which was observed to flare by INTEGRAL on 2004 October 8 (Rodriguez et al. 2004), and reach $\approx 70$ mCrab in the 20–60 keV energy range (integrated over 1700 s) and 20 mCrab in the 60–200 keV range. The IR counterpart is a B1 Ib star, 2MASS 18410043−0535465 (Halpern et al. 2004; Nespoli et al. 2008), and this firmly identifies it as a member of the SFXT class (Negueruela et al. 2006). Sguera et al. (2009) presented the first broad-band spectrum of this source for energies above 3 keV, obtained with *INTEGRAL* (IBIS+JEM-X).

AX J1841.0−0536 triggered BAT on 2010 June 5 at 17:23:30 UT (Romano et al. 2011b), and it was the first outburst of the source detected by the BAT for which *Swift* performed a slew, thus allowing broad-band data collection. Observations were performed both as automated target observations that lasted for several orbits, until $\sim 59$ ks after the trigger, and as follow-up target of opportunity (ToO) observations for a total of 10.8 ks. The XRT data cover the first 11 d after the beginning of the outburst. For the details of data reduction and analysis we refer to Romano et al. (2011b).

Figure 1a shows the first orbit of *Swift* data. The BAT light curves are rather flat and weak, but significant signal is found at the lower energies (15–50 keV). The XRT light curve (Figure 1b), on the contrary, is quite rich, starting from a maximum of $\sim 8$ counts s$^{-1}$, then decreasing to $\sim 0.01$ counts s$^{-1}$ during the first day, with several flares superimposed, hence yielding a dynamic range of approximately 900 during this outburst (Figure 1d). Then, after three days, the source count rate rose again and reached $\sim 1$ counts s$^{-1}$ (Figure 1b). We estimate that the observed dynamical range of this source in the XRT band (Figure 1c), considering the historical data we collected during our monitoring campaign (Sidoli et al. 2008; Romano et al. 2009) is $\sim 1600$, hence placing it well in the customary range for SFXTs.

A simple absorbed power-law model is inadequate in describing the broad band spectrum ($\chi^2_\nu = 1.6$ for 29 dof), so we considered models typically used to describe the X–ray emission from accreting pulsars in HMXBs, such as an absorbed power-law model with an exponential cutoff ($N_{\rm H} = 2.2^{+1.9}_{-1.1} \times 10^{22}$ cm$^{-2}$, $\Gamma = 0.2^{+0.7}_{-0.6}$, $E_{\rm c} = 16^{+21}_{-5}$ keV, $\chi^2_\nu/{\rm dof}= 1.2/28$) and an absorbed power-law model with a high energy cut-off ($N_{\rm H} = 1.9^{+1.7}_{-1.0} \times 10^{22}$ cm$^{-2}$, $\Gamma = 0.2^{+0.4}_{-0.5}$, $E_{\rm c} = 4^{+12}_{-4}$ keV, $E_{\rm f} = 16^{+10}_{-9}$ keV, $\chi^2_\nu/{\rm dof}= 1.2/27$). The latter model provide a more satisfactory fit of the broad-band emission, resulting in a hard powerlaw-like spectrum below 10 keV, with a roll over of the higher energies when simultaneous XRT and BAT data fits are performed. Figure 2 shows the fits for the absorbed power-law model with a high energy cut-off model. Furthermore, even though no statistically significant pulsations were found in the *Swift* data, AX J1841.0−0536 is one of the few SFXTs with a known pulse period (Bamba et al. 2001, $P_{\rm spin} = 4.7394 \pm 0.0008$ s). An indirect estimate of the magnetic field $B$ of the neutron star can be obtained from the HIGHECUT fit to the broad-band spectrum: our value of the high energy cutoff $E_{\rm c} < 16$ keV yields a $B \lesssim 3 \times 10^{12}$ G. This value for $B$, which is indeed similar to the one derived for the prototype of the SFXT class IGR J17544−2619, is inconsistent with a magnetar nature of AX J1841.0−0536.

In conclusion, AX J1841.0−0536 shows properties that resemble those of the prototype of the class, IGR J17544−2619, and fits well in the SFXT class, based on its observed properties during the 2010 outburst, its large dynamical range in X–ray luminosity, the similarity of the light curve (length and shape) to those of the other SFXTs observed by *Swift*, and the X–ray broad-band spectral properties.

## 3. Monitoring Programs

We have been performing several monitoring campaigns with *Swift*, starting from the outburst of the periodic SFXT IGR J11215−5952 in February 2007 (Romano et al. 2007), which demonstrated that the accretion during the bright outbursts lasts a few days, as opposed to a few hours, as observed by lower-sensitivity instruments.

It is worth mentioning the recent campaign on IGR J18483−0311 (Romano et al. 2010), during which we followed the X–ray light curve for 28 d, longer than a whole orbital period (18.52 d, Sguera et al. 2007). To constrain the different mechanisms proposed to explain the SFXT nature, we applied the new clumpy wind model for blue supergiants developed by Ducci et al.

Table 1. The *Swift* monitoring campaign and inactivity duty cycle. $\Delta T_\Sigma$ is sum of the exposures accumulated in all observations, each in excess of 900 s, where only a 3-$\sigma$ upper limit was achieved; $P_{\rm short}$ is the percentage of time lost to short observations; IDC is the duty cycle of inactivity; Rate$_{\Delta T_\Sigma}$ is the cumulative count rate (0.2–10 keV, in units of $\times 10^{-3}$ counts s$^{-1}$).

| Name | Campaign Dates | Obs. N. | Exposure (ks) | $\Delta T_\Sigma$ (ks) | $P_{\rm short}$ (%) | IDC (%) | Rate$_{\Delta T_\Sigma}$ ($10^{-3}$ counts s$^{-1}$) |
|---|---|---|---|---|---|---|---|
| IGR J16479–4514 | 2007-10-26–2009-11-01 | 144 | 161 | 29.7 | 3 | 19 | $3.1 \pm 0.5$ |
| XTE J1739–302   | 2007-10-27–2009-11-01 | 184 | 206 | 71.5 | 10 | 39 | $4.0 \pm 0.3$ |
| IGR J17544–2619 | 2007-10-28–2009-11-03 | 142 | 143 | 69.3 | 10 | 55 | $2.2 \pm 0.2$ |
| AX J1841.0–0536 | 2007-10-26–2008-11-15 | 88  | 96  | 26.6 | 3 | 28 | $2.4 \pm 0.4$ |

(2009) to the observed X–ray light curve and found that, assuming an eccentricity of $e = 0.4$, the X–ray emission from this source can be explained in terms of the accretion from a spherically symmetric clumpy wind, composed of clumps with different masses in the range (0.01–50) $\times 10^{20}$ g.

Between 2007 October 26 and 2009 November 3, we observed a sample of 4 SFXTs (IGR J16479−4514, XTE J1739−302/IGR J17391−3021, IGR J17544−2619, and AX J1841.0−0536/IGR J18410−0535) chosen among the 8 SFXTs known at the end of 2007. During the second year of *Swift* observations, we did not monitor AX J1841.0−0536. We obtained 2 or 3 XRT observations per week per source, each 1 ks long, to characterize their long-term behavior, to determine the properties of their quiescent state, to monitor the onset of the outbursts and to measure the outburst recurrence period and duration. During the two years of monitoring we collected 558 pointed XRT observations, for a total of 606 ks of on-source exposure. Table 1 summarizes the campaign. The long term X–ray properties outside the bright outbursts are described in Sidoli et al. (2008), Romano et al. (2009, 2011a), while the outbursts that occurred during the monitoring program are analyzed in detail in Romano et al. (2008), Sidoli et al. (2009a,b); Romano et al. (2011a).

### 3.1. Light curves and inactivity duty cycle

In Figure 3 we show the 0.2–10 keV XRT light curves for three of the four monitored SFXTs while in Figure 1c we show the fourth. We observe variability at all timescales and intensity ranges we can probe. As shown in Figure 4 for AX J1841.0−0536 [and in fig. 10 in Romano et al. (2011a) for the other three SFXTs], superimposed on the day-to-day variability is intra-day flaring which involves flux variations up to one order of magnitude that can occur down to timescales as short as $\sim 1$ ks, and which can be naturally explained by the accretion of single clumps composing the donor wind with masses of $M_{\rm cl} \sim 0.3$–$2 \times 10^{19}$ g.

Given this wealth of data, we can address the issue of the percentage of time each source spends in each flux state. Our monitoring can be considered a casual sampling of the light curves at a resolution of $\sim 3$–4 d over a $> 2$ yr baseline, therefore we estimate that these sources spend 3–5% of the total time in bright outbursts.

We considered the following states: BAT-detected outbursts; intermediate states (firm detections excluding outbursts); and 'non detections' (detections with a significance below $3\sigma$). From the latter state we excluded all observations that had a net exposure below 900 s [corresponding to 2–10 keV flux limits that vary between 1 and $3\times 10^{-12}$ erg cm$^{-2}$ s$^{-1}$ ($3\sigma$), depending on the source, see Romano et al. (2009)]. This was done because *Swift* is a GRB-chasing mission and several observations were interrupted by GRB events; therefore the consequent non detection may be due to the short exposure, and not exclusively to the source being faint. The duty cycle of *inactivity* is defined (see Romano et al. 2009, 2011a) as the time each source spends *undetected* down to a flux limit of 1–$3\times 10^{-12}$ erg cm$^{-2}$ s$^{-1}$,

$$\text{IDC} = \Delta T_\Sigma / [\Delta T_{\rm tot} (1 - P_{\rm short})], \quad (1)$$

where $\Delta T_\Sigma$ is sum of the exposures accumulated in all observations, each in excess of 900 s, where only a 3-$\sigma$ upper limit was achieved (Table 1, column 5), $\Delta T_{\rm tot}$ is the total exposure accumulated (Table 1, column 4), and $P_{\rm short}$ is the percentage of time lost to short observations (exposure $< 900$ s, Table 1, column 6). The cumulative count rate for each object is also reported Table 1 (column 8). We obtain that IDC = 19, 28, 39, 55 %, for IGR J16479−4514, AX J1841.0−0536, XTE J1739−302, IGR J17544−2619, respectively (Table 1, column 7), with an estimated error of $\sim 5$ %.

Based on the distributions of the observed count rates (after removal of the observations where a detection was not achieved), we find that the most probable 2–10 keV unabsorbed flux for these sources is $\sim 1$–$2 \times 10^{-11}$ erg cm$^{-2}$ s$^{-1}$ (Romano et al. 2011a), corresponding to luminosities in the order of a few $10^{33}$ to a few $10^{34}$ erg s$^{-1}$, two orders of magnitude lower than the bright outbursts, and still two orders of magnitude higher than the quiescent state.

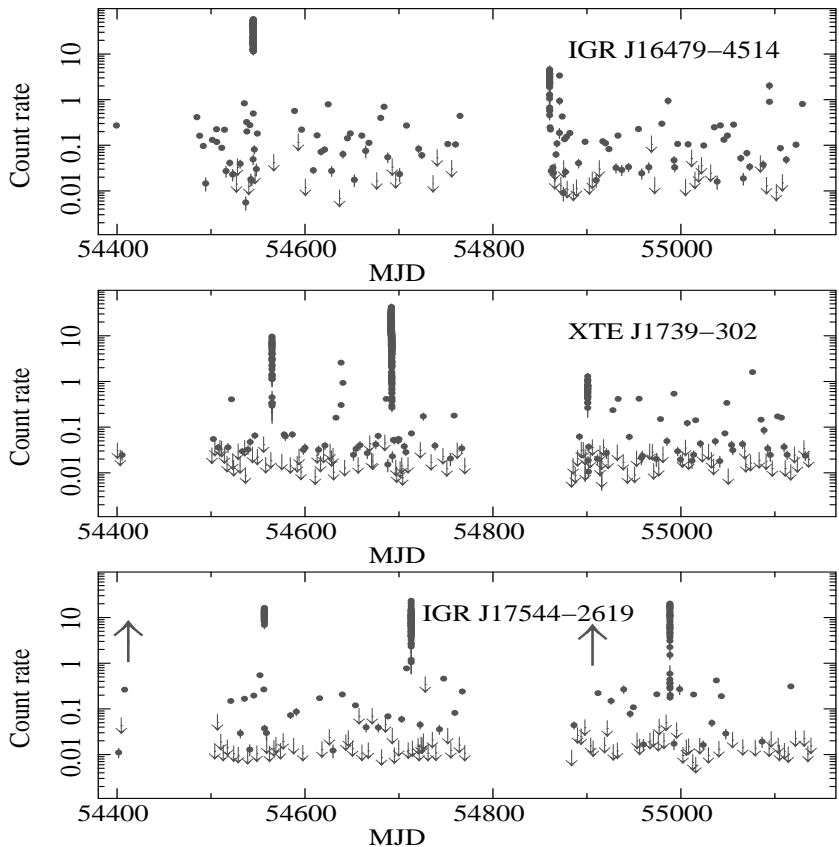

Fig. 3. *Swift*/XRT light curves in the 0.2–10 keV energy range, between 2007 October 26 and 2009 November 3. The light curves are background subtracted, corrected for pile-up (when required), PSF losses, and vignetting. Each point refers to the average flux observed during each observation performed with XRT, except for outbursts where the data were binned to include at least 20 source counts per time bin to best represent the dynamical range. Downward-pointing arrows are 3-$\sigma$ upper limits, upward pointing arrows mark either outbursts that XRT could not observe because the source was Sun-constrained, or BAT Transient Monitor bright flares. AX J1841.0−0536 was only observed during the first year and is reported in grey in Figure 1c. Adapted from Romano et al. (2011a).

From the spectral point of view, the campaigns have shown that fits performed in the 0.3–10 keV energy band by adopting simple models such as an absorbed power law or a blackbody (more complex models were not required by the data) result in hard power law photon indices (always in the range 0.8–2.1) or in hot blackbodies ($kT_{\rm BB} \sim 1$–2 keV).

4. Conclusions

Thanks to the *Swift* observations, the general picture we obtain is that, despite individual differences, common X–ray characteristics of this class are now well defined, such as outburst lengths well in excess of hours, with a multiple peaked structure, and a high dynamic range (including bright outbursts), up to $\sim 4$ orders of magnitude.

In the future we predict that MAXI, with its capability to monitor the whole sky in the 0.5–20 keV band, where the spectral energy distributions of SFXTs peak, and a a (5-$\sigma$) sensitivity of 60 mCrab in one orbit (15 mCrab in one day), will be able not only to observe bright outbursts of known SFXTs (such as the case here presented of AX J1841.0−0536) but also to discover new SFXTs. In both cases, the synergy between the *Swift* and MAXI observations will provide new insights in the SFXT phenomenon, as is currently being done for X–ray transients, such as MAXI J1659−152 (Kennea 2011a,b).

Acknowledgments
We acknowledge financial contribution from the agreement ASI-INAF I/009/10/0. This work was supported at PSU by NASA contract NAS5-00136. PE acknowledges financial support from the Autonomous Region of

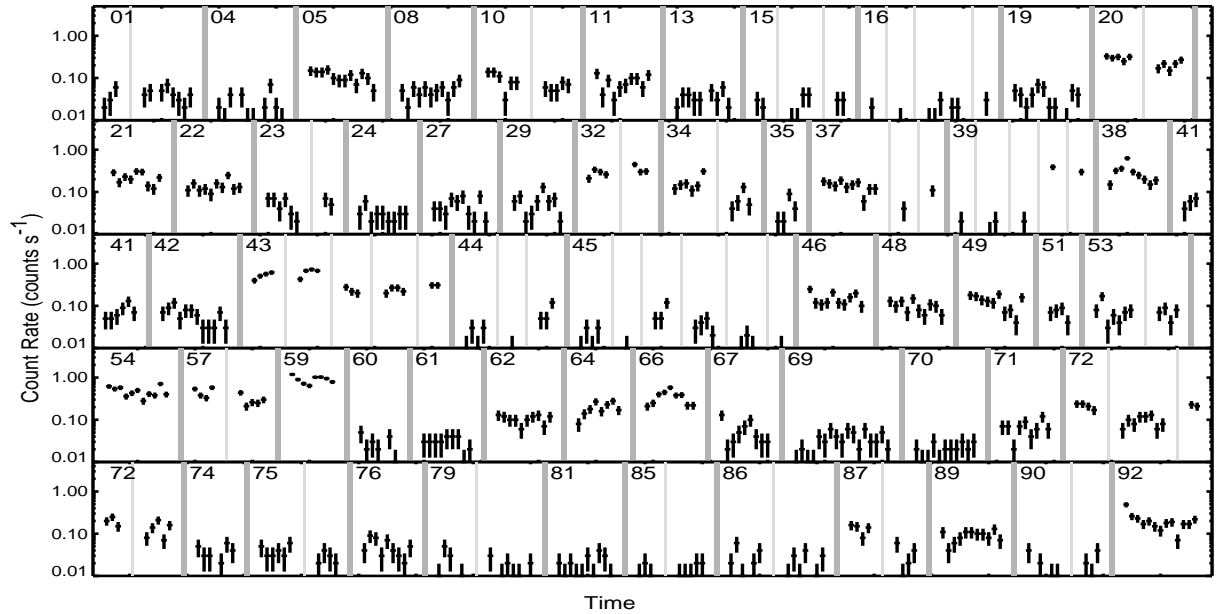

Fig. 4. *Swift*/XRT (0.2–10 keV) montage of time sequences in the monitoring campaign on AX J1841.0–0536. The numbers in the plot identify each observing sequence (see Table 8 in Romano et al. 2009). Non observing intervals and orbital gaps have been cut out from the time axis and replaced by thick vertical bars to separate different sequences and thin grey bars to separate different orbits within each sequence. Each point represents a 100 s bin.

Sardinia through a research grant under the program PO Sardegna FSE 2007–2013, L.R. 7/2007 "Promoting scientific research and innovation technology in Sardinia".

References

Bamba A. et al. 2001 PASJ, 53, 1179
Barthelmy S. D. et al. 2005 Space Science Rev., 120, 143
Bozzo E., Falanga M., Stella L. 2008 ApJ, 683, 1031
Burrows D. N. et al. 2005 Space Science Rev., 120, 165
Ducci L. et al. 2009 MNRAS, 398, 2152
Gehrels N. et al. 2004 ApJ, 611, 1005
Grebenev S. A., Sunyaev R. A., 2007 Astronomy Letters, 33, 149
Halpern J. P., Gotthelf E. V., 2004 Astron. Tel., 341
Halpern J. P. et al. 2004 Astron. Tel., 289
in't Zand J. J. M. 2005 A&A, 441, L1
Kennea J. A. 2011a ApJ, in press
Kennea J. A. 2011b These Proceedings
Matsuoka M. et al. 2009 PASJ, 61, 999
Negoro H. et al. 2010 Astron. Tel., 3018
Negueruela I et al. 2006 ESA SP 604, 165
Negueruela I. et l. 2008 1010, 252
Nespoli E. et l. 2008 A&A, 486, 911
Rodriguez J. et al. 2004 Astron. Tel., 340
Romano P. et al. 2007 A&A, 469, L5
Romano P. et al. 2008 ApJL, 680, L137
Romano P. et al. 2009 MNRAS, 399, 2021
Romano P. et al. 2010 MNRAS, 401, 1564
Romano P. et al. 2011a MNRAS, 410, 1825
Romano P. et al. 2011b MNRAS Letters, in press, arXiv:1012.0028
Sguera V. et al. 2005 A&A, 444, 221
Sguera V. et al. 2006 ApJ, 646, 452
Sguera V. et al. 2007 A&A, 467, 249
Sguera V. et al. 2009 ApJ, 697, 1194
Sidoli L. et al. 2006 A&A, 450, L9
Sidoli L. et al. 2007 A&A, 476, 1307
Sidoli L. et al. 2008 ApJ, 687, 1230
Sidoli L. et al. 2009a ApJ, 690, 120
Sidoli L. et al. 2009b MNRAS, 397, 1528
Walter R., Zurita Heras J., 2007 A&A, 476, 335
Walter R. et al. 2006 A&A, 453, 133